\pdfoutput=1
\documentclass[conference]{IEEEtran}
\IEEEoverridecommandlockouts
\usepackage[utf8]{inputenc}
\usepackage{cite}
\usepackage[letterpaper, left=0.625in, right=0.625in, bottom=1in, top=0.75in]{geometry}
\usepackage{amsmath,amssymb,amsfonts}
\usepackage{algorithm} 
\usepackage{algpseudocode}
\usepackage{graphicx}
\usepackage{caption}
\usepackage{subcaption}
\usepackage{textcomp}
\usepackage{xcolor}
\usepackage{amsthm, bm, mathrsfs, indentfirst, longtable}
\usepackage[bookmarks=false,hidelinks]{hyperref}
\usepackage{longtable}
\usepackage{tabularx,booktabs,makecell,array,siunitx}
\usepackage{booktabs,float}
\usepackage{bbm}
\usepackage{balance}
\usepackage{threeparttable}
\usepackage{multirow}
\usepackage{booktabs}
\usepackage{lipsum}

\allowdisplaybreaks
\makeatletter
\renewcommand{\IEEEauthorblockA}[1]{
  \begingroup
  \normalsize #1\par
  \endgroup}
\makeatother
\usepackage{amsfonts}
\begin{document}

\title{Segment-Wise Flow Matching for Vision-Aided mmWave V2I Beam Prediction}

\author{
\linespread{1.0}\selectfont 
\IEEEauthorblockN{Can Zheng\IEEEauthorrefmark{1}\IEEEauthorrefmark{2}, Jiguang He\IEEEauthorrefmark{2}, Chung G. Kang\IEEEauthorrefmark{1}, Guofa Cai\IEEEauthorrefmark{3}, Chongwen Huang\IEEEauthorrefmark{4}, Henk Wymeersch\IEEEauthorrefmark{5}}\\
\IEEEauthorblockA{\IEEEauthorrefmark{1}School of Electrical Engineering, Korea University, Seoul, Republic of Korea}\\
\IEEEauthorblockA{\IEEEauthorrefmark{2}School of Computing and Information Technology, Great Bay University, Dongguan 523000, China}\\
\IEEEauthorblockA{\IEEEauthorrefmark{3}School of Information Engineering, Guangdong University of Technology, Guangzhou, China}\\
\IEEEauthorblockA{\IEEEauthorrefmark{4}College of Information Science and Electronic Engineering, Zhejiang University, Hangzhou 310027, China}\\
\IEEEauthorblockA{\IEEEauthorrefmark{5}Department of Electrical Engineering, Chalmers University of Technology, Gothenburg, Sweden}
}

\maketitle
\begin{abstract}
    This paper proposes a vision-conditioned flow matching (FM) framework for beam prediction in millimeter-wave vehicle-to-infrastructure links. Instead of modeling discrete beam-index sequences, the proposed method \textcolor{black}{learns local beam evolution from interpolations between adjacent normalized beam receive power vectors} through a continuous vector field governed by an ordinary differential equation, enabling smooth dynamics and efficient sampling. By imposing FM on \textcolor{black}{beam evolution between adjacent time steps} and jointly optimizing beam prediction and flow consistency, the proposed framework provides a unified model for future beam prediction. Experimental results show that the proposed FM-based model significantly improves beam prediction performance over baselines, approaches the performance of large language model–based methods, and reduces predictor-side inference latency by about $6.9\times$ on GPU and $2.8\times10^3\times$ on CPU, respectively.
\end{abstract}


\section{Introduction}

    Millimeter-wave (mmWave) vehicle-to-infrastructure (V2I) links offer high throughput but suffer from frequent misalignment and blockage under mobility, making repeated beam training and sweeping costly in urban scenarios \cite{3GPP38900, DLBM}. Sensing-aided beam prediction mitigates this by using perceptual data, such as visual inputs from roadside cameras, to predict optimal beams without full channel state information (CSI) \cite{vision}. \textcolor{black}{However, maintaining reliable long-horizon prediction remains challenging for traditional sequence predictors (e.g., RNNs and LSTMs) \cite{CV},} while recent multimodal and Transformer-based methods require large paired datasets and heavy computation \cite{transformer, BeamLLM, CV+GPS}, limiting real-time deployment at resource-constrained roadside units (RSUs). These limitations call for a modeling paradigm that treats beam evolution as a smooth continuous process and supports accurate, low-latency prediction with lightweight inference.

    Flow matching (FM) offers exactly such a paradigm: it has recently emerged as a framework for learning continuous-time dynamics that combine high predictive fidelity with efficient sampling \cite{FM}. By modeling data evolution as a continuous conditional flow governed by ordinary differential equations (ODEs), FM enables fast and stable sampling with only a few integration steps.
    It has achieved competitive or faster generation across images, videos, speech, and molecular domains \cite{FM,FM_video,FM_speech}, and has recently been adapted to autonomous driving for trajectory generation and planning \cite{FM_Auto}, showing its suitability for tasks with smoothly evolving states and stringent fast-sampling requirements.

    Inspired by the effectiveness of FM, we design a vision-conditioned beam prediction framework. The main contributions of this paper are summarized as follows:
    (i) \textbf{Flow-based modeling in the beam-codebook power domain:} We introduce a vision-conditioned FM framework that \textcolor{black}{learns local beam evolution from adjacent normalized beam receive power vectors} over the predefined codebook, rather than directly predicting discrete beam index sequences;
    (ii) \textbf{Segment-wise FM for temporal evolution:} We impose FM over \textcolor{black}{adjacent power vectors}, allowing the model to learn \textcolor{black}{local vector fields} that better capture the step-by-step temporal evolution of beam dynamics; and
    (iii) \textbf{Efficient and practical edge deployment:} The proposed FM model achieves superior prediction accuracy with lightweight computational cost, making it a promising lightweight predictor for edge-side beam prediction.
    \begin{figure}[t]
    \captionsetup{font=footnotesize}
    \begin{center}
    \includegraphics[width=\linewidth]{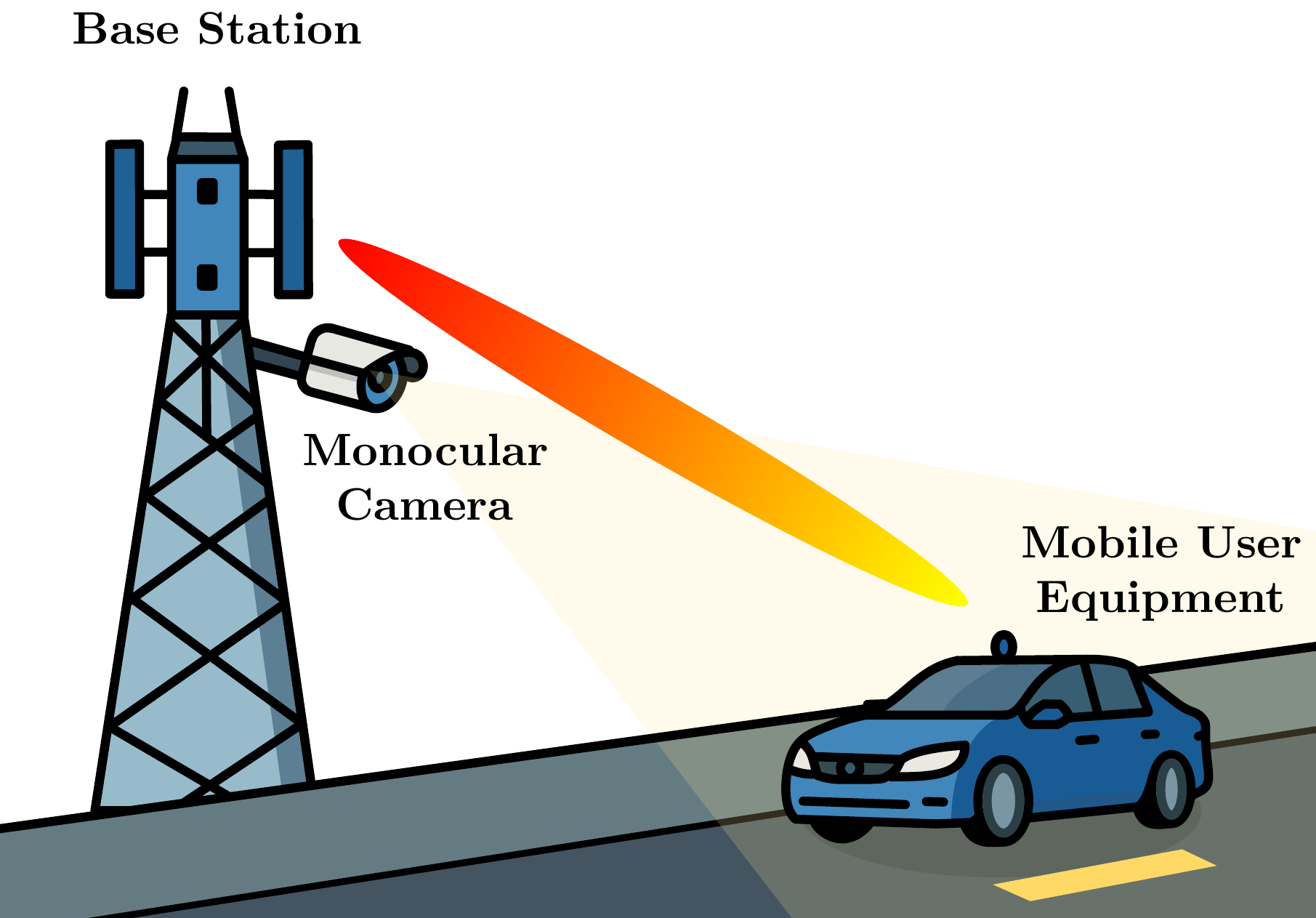}
    \end{center}
    \caption{Illustration of the V2I system model, where the RSU is equipped with a monocular camera.}
    \label{fig:V2I}
    \end{figure}

\textit{Notations: }
Bold lowercase and uppercase letters denote vectors, matrices, and tensors. The superscripts $(\cdot)^{\mathsf{T}}$ and $(\cdot)^{\mathsf{H}}$ represent the transpose and Hermitian operations, respectively.
We define a sequence as $\mathbf{X}_{a:b}\triangleq\{\mathbf{x}_\tau\}_{\tau=a}^{b}$. We denote by $\mathbf{x}_a$ the $a$-th element of the vector $\mathbf{x}$. Standard norms include the Euclidean norm $|\cdot|_2$ and the element-wise magnitude $|\cdot|$. The primary operators used are the expectation $\mathbb{E}[\cdot]$, the indicator function $\mathbbm{1}\{\cdot\}$, and the maximizer index $\arg\max(\cdot)$. The $\mathrm{SOFTMAX}(\cdot)$ function maps logits to probabilities. Furthermore, $\dot{\mathbf{z}}_\tau = d\mathbf{z}_\tau/d\tau$ is the derivative with respect to \textcolor{black}{the local interpolation parameter $\tau$}, and the standard fields $\mathbb{R}$ and $\mathbb{C}$ denote the real and complex numbers.

\section{System Model}
\label{chapt2}    
    \subsection{System Description}
    \label{IIA}
    
    We consider a roadside mmWave system where an RSU is equipped with an $N$-element antenna array and employs a single radio frequency (RF) chain and a phase shifter-based analog beamformer. This RSU serves a single-antenna user equipment (UE), as illustrated in Fig.~\ref{fig:V2I}. The RSU applies analog beamforming using a predefined codebook $\mathcal{W}=\{\mathbf{w}_1,\dots,\mathbf{w}_M\}$, where $M$ is the size of the codebook, and each $\mathbf{w}_m\in\mathbb{C}^{N}$ is unit norm. The downlink narrowband baseband model at time $t$ is given by
    \begin{align}
        r_t = \mathbf{h}_t^{\mathsf H}\mathbf{w}_m s_t + n_t,
    \end{align}
    where $\mathbf{h}_t\in\mathbb{C}^{N}$ is the downlink channel vector, $s_t$ is a unit-power symbol, and $n_t$ is complex Gaussian noise.
    
    For the predefined beam codebook, we define the beam receive-power vector at time $t$ as
    \begin{align}
        \mathbf{p}_t =
        \left[
        \big|\mathbf{h}_t^{\mathsf H}\mathbf{w}_1\big|^2,\,
        \big|\mathbf{h}_t^{\mathsf H}\mathbf{w}_2\big|^2,\,
        \dots,\,
        \big|\mathbf{h}_t^{\mathsf H}\mathbf{w}_M\big|^2
        \right]^{\mathsf T}
        \in \mathbb{R}^{M}.
    \end{align}
    Accordingly, the optimal beam index that maximizes the instantaneous receive power is
    \begin{align}
        m_t^{*}=\arg\max_{m\in\{1,\dots,M\}} \mathbf{p}_{t,m}.
    \end{align}

    \textcolor{black}{To provide the visual observations used for beam prediction, a} monocular RGB camera mounted on the RSU captures the roadway ahead. From each frame, a pre-trained object detector extracts the bounding box \textcolor{black}{(bbox)} of the target vehicle, denoted as $\mathbf{x}_t = [x_{c,t},\,y_{c,t},\,w_t,\,h_t]^{\mathsf T}$, where $(x_{c,t},y_{c,t})$ is the box center and $(w_t,h_t)$ are its width and height of the box.

    \subsection{Objective and Challenge}

    We define $T_{\mathrm{Hist}}$ and $T_{\mathrm{Pred}}$ as the history length and prediction horizon (both in frames), respectively. 
    For a reference time $t$, the RSU observes the past visual history of length $T_{\mathrm{Hist}}$ and aims to predict the optimal beam indices for the current and future $T_{\mathrm{Pred}}$ time steps. We denote this mapping as
    \begin{align}
        \hat{\mathbf{Y}}_{t:t+T_{\mathrm{Pred}}-1}
        = f_{\Theta}\big(\mathbf{X}_{t-T_{\mathrm{Hist}}:t-1}\big),
    \end{align}
    where $f_{\Theta}(\cdot)$ is a learned predictor parameterized by $\Theta$,
    $\mathbf{X}_{t-T_{\mathrm{Hist}}:t-1}$ is the visual input sequence\footnote{\textcolor{black}{Throughout this paper, ``visual input'' refer to camera-derived bbox sequences rather than raw RGB pixels.}}, and the model output
    $\hat{\mathbf{Y}}_{t:t+T_{\mathrm{Pred}}-1}$ is a sequence, where for each time step $t' \in \{t, t+1, \dots, t+T_{\mathrm{Pred}}-1\}$, each column $\hat{\mathbf{y}}_{t'}\in \mathbb{R}^{M}$ is a probability distribution vector over the available beams.
    The resulting predicted beam index sequence $\hat{m}_{t:t+T_{\mathrm{Pred}}-1}$ is derived by taking the index corresponding to the maximum probability element in each $\hat{\mathbf{y}}_{t'}$, and this sequence targets the ground-truth indices $m_{t:t+T_{\mathrm{Pred}}-1}^{*}$ defined in
    Section~\ref{IIA}. 
    
    This vision-conditioned beam prediction task is challenging for several reasons: First, the mapping from 2D \textcolor{black}{bbox} trajectories to the future optimal beam indexes is highly complex and environment-dependent. Specifically, the underlying wireless channel evolution, and thus the mapping from visual state to optimal beam index, is fundamentally a non-linear, non-Gaussian, and non-Markovian process. Even though the channel is largely geometric, the inherent time-varying nature of V2I environments introduces channel complexity. In addition, the high sensitivity of narrow beams to small angular changes results in a prediction output highly sensitive to the input's spatio-temporal dynamics, making traditional methods inadequate. To tackle these challenges, we utilize a historical sensing data sequence for spatio-temporal prediction to capture the UE's motion trend and the dynamic environmental context, which also aligns with the requirements for beam prediction outlined in 3rd generation partnership project (3GPP) artificial intelligence (AI)/machine learning (ML) for beam management (BM) Cases 1 and 2 \cite{3gpp}. \textcolor{black}{Second, the predictor should capture the coherent temporal evolution of the underlying beam power vectors across the multi-step prediction horizon, which becomes increasingly challenging as the prediction horizon grows \cite{error}.} Third, for high-mobility V2I scenarios, prediction must have extremely low latency to enable beam switching before channel conditions change \cite{BM}. This requires the inference process to remain lightweight for deployment on resource-constrained RSUs.
    These considerations motivate a design that (i) models smooth beam dynamics in a continuous latent space, \textcolor{black}{(ii) jointly models beam power evolution over a finite prediction horizon,} and (iii) enables fast, stable inference. In the next section, we present a flow-matching-based beam prediction framework that meets these requirements.
    
\section{Vision-Aided Beam Prediction via Flow Matching}
\label{chapt3}
    \begin{figure}[t]
    \captionsetup{font=footnotesize}
    \begin{center}
    \includegraphics[width=\linewidth]{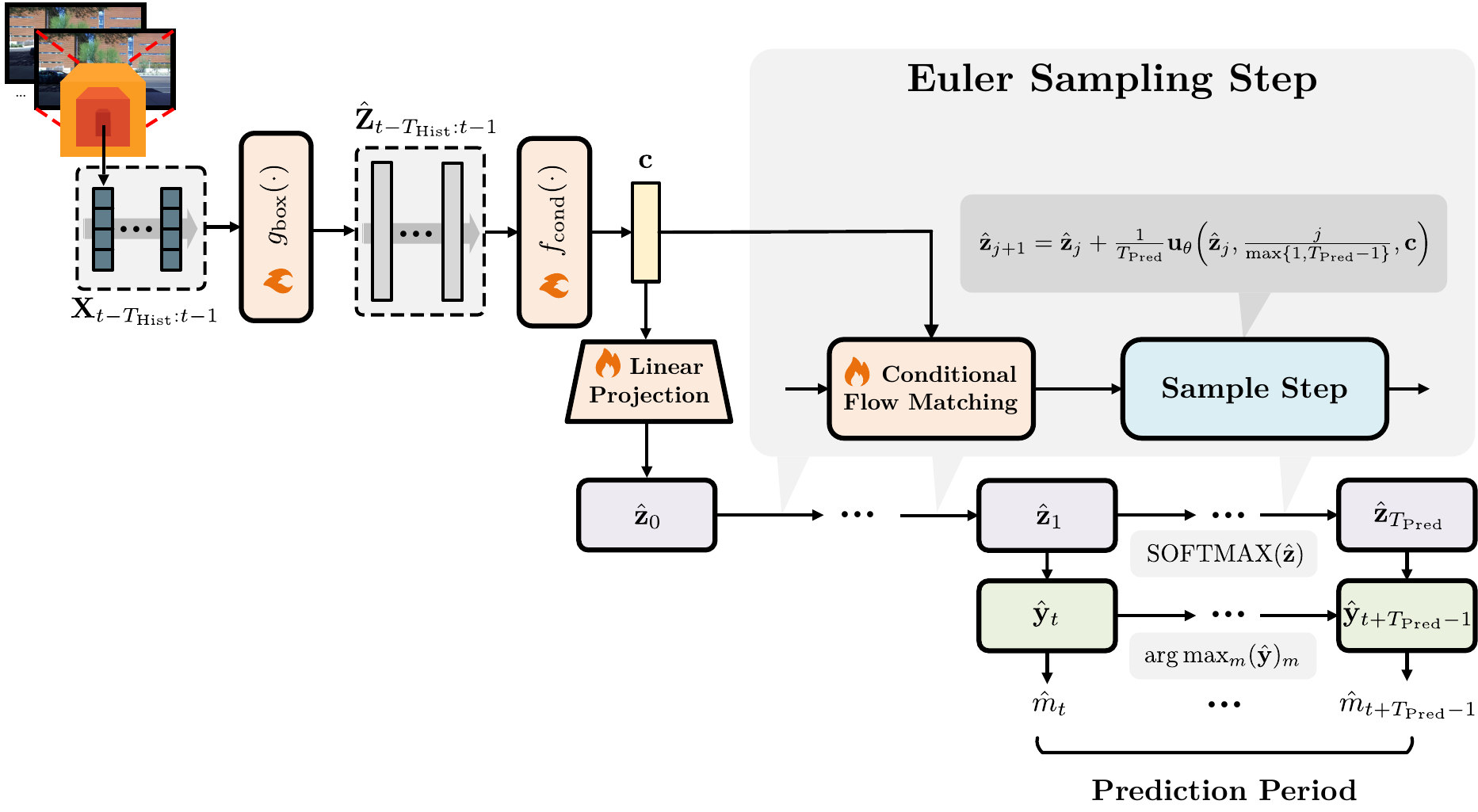}
    \end{center}
    \caption{Architecture of the proposed FM-based beam prediction framework. Modules with a spark icon participate in end-to-end training.}
    \label{fig:FM}
    \end{figure}

    Fig.~\ref{fig:FM} illustrates the proposed FM-based beam prediction framework, which aims to learn a continuous conditional flow in the latent space, enabling \textcolor{black}{smooth beam evolution over time}. The proposed framework comprises an FM training objective, a vision-conditioned beam flow model, and \textcolor{black}{an Euler solver for efficient inference}.

    \subsection{FM Training}
    We model the temporal dynamics of receive powers over the predefined codebook $\mathcal{W}$. To this end, we represent each time step by the normalized beam receive power vector as
    \begin{align}
        \tilde{\mathbf{p}}_t = \mathrm{SOFTMAX}(\mathbf{p}_t) \in \mathbb{R}^{M},
    \end{align}
    where the $m$-th element of $\tilde{\mathbf{p}}_t$ represents the normalized receive power associated with the $m$-th beam.

    \textcolor{black}{For each adjacent segment index $t'$, we independently sample a local interpolation parameter $\tau\sim\mathcal{U}(0,1)$. The corresponding vector $\mathbf{z}_\tau\in\mathbb{R}^M$ satisfies}
    \begin{align}
        \dot{\mathbf{z}}_\tau = \mathbf{u}_\theta(\mathbf{z}_\tau,\tau),
        \label{eq:flow}
    \end{align}
    where $\mathbf{u}_\theta(\cdot,\tau)$ is a learnable vector field parameterized by $\theta$.
    
    We construct a local linear interpolant between \textcolor{black}{adjacent normalized beam power vectors $\tilde{\mathbf{p}}_{t'}$ and $\tilde{\mathbf{p}}_{t'+1}$ as}
    \begin{align}
        \mathbf{z}_\tau=\textcolor{black}{(1-\tau)\tilde{\mathbf{p}}_{t'}+\tau\tilde{\mathbf{p}}_{t'+1}},
        \label{eq:z_tau}
    \end{align}
    \begin{align}
        \dot{\mathbf{z}}_\tau=\textcolor{black}{\tilde{\mathbf{p}}_{t'+1}-\tilde{\mathbf{p}}_{t'}}.
        \label{eq:dot_z_tau}
    \end{align}
    \textcolor{black}{This gives an analytic target vector field $\dot{\mathbf{z}}_\tau$, defined as the derivative of the interpolated vector with respect to the local parameter $\tau$, to supervise $\mathbf{u}_\theta$.} The use of local FM is justified by the extremely short duration of each adjacent temporal segment.\footnote{The data sampling rate of the dataset considered in this work is approximately 6--8 frames per second (FPS). \cite{deepsense6g}.} \textcolor{black}{Within such a short time frame, beam evolution between two adjacent time steps can be reasonably approximated by a linear path between the corresponding normalized power vectors, yielding a piecewise-constant target vector field within each segment.}
    
    Accordingly, we define the FM loss as the \textcolor{black}{following squared difference between the target vector field and the learned vector field}:
    \begin{align}
        \mathcal{L}_\mathrm{FM}(\theta)
        = \frac{1}{M T_{\mathrm{Pred}}}
        \sum_{t'=t-1}^{t+T_{\mathrm{Pred}}-2}
        \mathbb{E}_{\tau} 
        \left[
        \left\Vert
        \mathbf{u}_\theta (\mathbf{z}_\tau, \tau)
        -\dot{\mathbf{z}}_\tau
        \right\Vert_2^2
        \right].
    \label{eq:FM_loss}
    \end{align}
    Minimizing $\mathcal{L}_{\mathrm{FM}}$ encourages $\mathbf{u}_\theta$ to approximate the local temporal derivative of the beam receive power vector, so the learned vector field captures the step-by-step evolution of beam powers over consecutive frames.

    \begin{algorithm}[t]
    \caption{Vision-conditioned FM (training and inference)}
    \label{alg:joint_fm}
    \begin{algorithmic}[1]
    \Require Codebook size $M$; history $T_{\mathrm{Hist}}$; horizon $T_{\mathrm{Pred}}$; input visual data $\mathbf{X}_{t-T_{\mathrm{Hist}}:t-1}$.
    \Statex \textbf{// Training Objectives:} Box embedding function $g_{\mathrm{box}}(\cdot)$, condition encoder $f_{\mathrm{cond}}(\cdot)$, \textcolor{black}{learnable linear initialization projection, and} conditional vector field $\mathbf{u}_\theta$\textcolor{black}{.}
    \Statex \textbf{// Training Phase}
    \For{each minibatch $\mathcal{B}$}
      \State Obtain the condition vector $\mathbf{c}$ according to \eqref{eq:condi}
      \Statex \textbf{FM Loss}
      \State \textcolor{black}{Initialize the accumulated FM discrepancy}
      \For{$t'=t-1$ \textbf{to} $t{+}T_{\mathrm{Pred}}{-}2$}
        \State \textcolor{black}{Independently sample $\tau\sim\mathcal{U}(0,1)$}
        \State \textcolor{black}{Construct $\mathbf{z}_\tau$ and $\dot{\mathbf{z}}_\tau$ according to~\eqref{eq:z_tau} and~\eqref{eq:dot_z_tau}}
        \State \textcolor{black}{Accumulate the squared discrepancy using~\eqref{eq:cond_ODE}}
      \EndFor
      \State \textcolor{black}{Average the discrepancy according to~\eqref{eq:FM_loss}}
      \Statex \textbf{Classification Loss}
      \State \textcolor{black}{Obtain $\hat{\mathbf{z}}_{0}$ from $\mathbf{c}$ through a learnable linear projection}
      \For{$j=0$ \textbf{to} $T_{\mathrm{Pred}}{-}1$}
        \State \textcolor{black}{Compute $\hat{\mathbf{z}}_{j+1}$ using the Euler integration in~\eqref{eq:update}}
      \EndFor
      \State \textcolor{black}{Obtain $\hat{\mathbf{y}}_{t+j}$ from $\hat{\mathbf{z}}_{j+1}$ according to~\eqref{eq:y}, for all $j=0,\ldots,T_{\mathrm{Pred}}-1$}
      \State Compute $\mathcal{L}_{\mathrm{CE}} $ according to~\eqref{eq:CE_loss}
      \State \textbf{Update} $\Theta$ to minimize  $\mathcal{L}$
    \EndFor
    \Statex \textbf{// Inference Phase}
    \State Obtain the condition vector $\mathbf{c}$ according to \eqref{eq:condi}
    \State \textcolor{black}{Obtain $\hat{\mathbf{z}}_{0}$ from $\mathbf{c}$ through a learnable linear projection}
    \For{$j=0$ \textbf{to} $T_{\mathrm{Pred}}{-}1$}
      \State \textcolor{black}{Compute $\hat{\mathbf{z}}_{j+1}$ using the Euler integration in~\eqref{eq:update}}
    \EndFor
    \State \textcolor{black}{Obtain $\hat{\mathbf{y}}_{t+j}$ from $\hat{\mathbf{z}}_{j+1}$ according to~\eqref{eq:y}, $\forall j$}
    \State \textbf{Return} \textcolor{black}{Predicted beams $\hat{m}_{t+j}=\arg\max_m\hat{\mathbf{y}}_{t+j,m}$, $\forall j$}.
    \end{algorithmic}
    \end{algorithm}

    \subsection{Vision-Conditioned Beam Flow}
    In realistic vehicular scenes, beam evolution depends strongly on visual context. We first use a learnable embedding function $g_{\mathrm{box}}(\cdot)$ to transform each \textcolor{black}{bbox} vector $\mathbf{x}_{t'} \in \mathbb{R}^4$ into a \textcolor{black}{bbox feature vector in $\mathbb{R}^{4M}$}. The historical feature sequence $\hat{\mathbf{Z}}_{t-T_{\mathrm{Hist}}:t-1} = g_{\mathrm{box}}(\mathbf{X}_{t-T_{\mathrm{Hist}}:t-1})$ is then input to a temporal--spatial condition encoder $f_{\mathrm{cond}}(\cdot)$ to extract the context feature $\mathbf{c}$:
    \begin{align}
        \mathbf{c} &= f_{\mathrm{cond}}\left(\hat{\mathbf{Z}}_{t-T_{\mathrm{Hist}}:t-1}\right) \in \mathbb{R}^{d_c}.\label{eq:condi}
    \end{align}
    
    Conditioning on $\mathbf{c}$ \textcolor{black}{gives}
    \begin{align} 
       \dot{\mathbf{z}}_\tau= \mathbf{u}_\theta({\mathbf{z}}_\tau, \tau, \mathbf{c}),\quad \tau \in [0,1].
       \label{eq:cond_ODE}
    \end{align}
    \textcolor{black}{The initial vector $\hat{\mathbf{z}}_0\in\mathbb{R}^M$ is obtained from $\mathbf{c}$ through a learnable linear projection. Euler integration is then performed as follows:}
    \begin{align}
        \hat{\mathbf{z}}_{j+1}
        &={\color{black}\hat{\mathbf{z}}_j+\frac{1}{T_{\mathrm{Pred}}}
        \mathbf{u}_{\theta}\!\left(\hat{\mathbf{z}}_j,\frac{j}{\max\{1,T_{\mathrm{Pred}}-1\}},\mathbf{c}\right)},
        \label{eq:update}
    \end{align}
    \textcolor{black}{for $j=0,\ldots,T_{\mathrm{Pred}}-1$. This Euler integration produces $T_{\mathrm{Pred}}$ predicted vectors $\hat{\mathbf{z}}_1,\ldots,\hat{\mathbf{z}}_{T_{\mathrm{Pred}}}$ over the prediction horizon.} In this way, the learned conditional vector field captures how the beam receive power evolves over the predefined codebook under visual context, \textcolor{black}{thereby enabling future beam prediction}. \textcolor{black}{The interpolation in~\eqref{eq:z_tau} is used only to construct the local path for FM training. During prediction, $\hat{\mathbf{z}}_j$ is not required to be normalized.}

    Each \textcolor{black}{predicted vector is converted into beam probabilities via softmax}:
    \begin{align}
        \hat{\mathbf{y}}_{t+j} = \textcolor{black}{\mathrm{SOFTMAX}\big(\hat{\mathbf{z}}_{j+1}\big),\quad j=0,\ldots,T_{\mathrm{Pred}}-1.}
        \label{eq:y}
    \end{align}
    The cross-entropy loss is then computed as
    \begin{align}
        \mathcal{L}_{\mathrm{CE}}
        = - \frac{1}{T_\mathrm{Pred}}\sum_{t'=t}^{t+T_{\mathrm{Pred}}-1}
            \log (\hat{\mathbf{y}}_{t',{m_{t'}^{*}}}),
        \label{eq:CE_loss}
    \end{align}
    where ${(\hat{\mathbf{y}}_{t'})}_{m_{t'}}$ is the predicted probability of selecting beam $m$ at time $t'$.

    The final objective combines local flow alignment and classification terms. To balance the classification and FM tasks adaptively during training, we adopt the uncertainty weighting strategy \cite{training}. The final training objective is written as
    \begin{align}
        \mathcal{L}
        &= e^{-s_{\mathrm{CE}}}\,\mathcal{L}_{\mathrm{CE}} + e^{-s_{\mathrm{FM}}}\,\mathcal{L}_{\mathrm{FM}} + \frac{1}{2}(s_{\mathrm{CE}}+s_{\mathrm{FM}}),
    \end{align}
    where $s_{\mathrm{CE}}$ and $s_{\mathrm{FM}}$ are learnable log-variance parameters associated with the classification and FM tasks, respectively.
    
    \subsection{Inference Process}
    At inference, the condition vector $\mathbf{c}$ is derived from the observed visual history. Specifically, the \textcolor{black}{bbox} observations $\mathbf{X}_{t-T_{\mathrm{Hist}}:t-1}$ are first transformed into a feature sequence according to~\eqref{eq:condi}. \textcolor{black}{The initial vector $\hat{\mathbf{z}}_0$ is obtained from $\mathbf{c}$ through the learnable linear projection, and the Euler solver in~\eqref{eq:update} produces $T_{\mathrm{Pred}}$ predicted vectors. Each predicted vector is converted into beam probabilities according to~\eqref{eq:y},} and the predicted index is \textcolor{black}{$\hat{m}_{t+j}=\arg\max_m\hat{\mathbf{y}}_{t+j,m}$}.
        
        For clarity, the training and inference procedures are summarized in \textbf{Algorithm~\ref{alg:joint_fm}}.

\section{Simulation Results}
\label{chapt4}
    \subsection{Scenario, Methods, and Metrics}
    We use Scenario 8 of the DeepSense 6G dataset for performance evaluation \cite{deepsense6g}, which features a relatively confined region of interest (ROI) and predominantly LoS-dominant conditions. \textcolor{black}{Bboxes} of target users and candidate users are extracted from the camera images using YOLOv4 \cite{yolov4}. As the focus of this work is the modeling of beam evolution, we consider a single-user beam prediction setting and assume ideal target user identification when multiple vehicles appear in the camera view \cite{multicandidate}.
    Supervised samples are constructed with a fixed window length $T=13$, split into historical and prediction segments $(T_\mathrm{Hist},T_\mathrm{Pred})$. For each anchor time $t$, the model predicts the future beam sequence $\{m^{*}_{t:t+T_\mathrm{Pred}-1}\}$. 
    To prevent data leakage, sequence-level random partitioning is adopted before window sampling.
    We evaluate two configurations: \textbf{configuration A} ($T_\mathrm{Hist}/T_\mathrm{Pred}=8/5$) and \textbf{configuration B} ($3/10$). \textcolor{black}{The default training settings and model architecture are summarized in Tables~\ref{tab:param} and~\ref{tab:hypers}, respectively.}
    
\begin{table}[t]
    \centering
    \caption{Default training settings.}
    \label{tab:param}
    \begin{tabular}{ll}
    \toprule
    \textbf{Parameter} & \textbf{Value} \\
    \midrule
    Batch size & 64 \\
    Codebook size ($M$) & 32 \\
    Training epochs & 100 \\
    Optimizer & AdamW \\
    Initial learning rate & $10^{-3}$ \\
    Learning-rate schedule & 10\% warmup + 90\% cosine decay \\
    \bottomrule
    \end{tabular}
    \end{table}

    \begin{table}[t]
        \centering
        \caption{Model architecture summary.}
        \label{tab:hypers}
        \begin{tabular}{lll}
            \toprule
            \textbf{Module} & \textbf{Configurations} & \textbf{Dim.} \\
            \midrule
            $g_\mathrm{box}$ & MLP $4 \rightarrow 4M \rightarrow 4M$ & $4M$ \\
            $f_\mathrm{cond}$ & $1$-layer Transformer encoder & $4M$ \\
            $\mathbf{u}_\theta$ & MLP on $[\mathbf{z}, \tau, \mathbf{c}]$ & $M$ \\
            \bottomrule
        \end{tabular}
        \begin{tablenotes}
            \footnotesize
            \item \textit{Note:} ``Dim.'' denotes the output dimension, and ``MLP'' denotes multilayer perceptron.
        \end{tablenotes}
    \end{table}

    We compare against classical sequence models (RNN and long short-term memory, LSTM) trained on the same visual inputs, \textcolor{black}{i.e., the same target bbox sequences $\mathbf{X}_{t-T_{\mathrm{Hist}}:t-1}$.} We also include BeamLLM~\cite{BeamLLM} as a recent large-model baseline. Ablations isolate the contribution of each objective ($\mathcal{L}_{\mathrm{FM}}$, $\mathcal{L}_{\mathrm{CE}}$) and compare different condition encoders (Transformer versus RNN and LSTM).

    Let ${m^{(n)}_{t'}}^{*}$ denote the index of the optimal beam (ground-truth label) for the $n$-th test sample at time $t'$.  
    We compute the top-$K$ accuracy at time $t'$ by averaging over all test samples:
    \begin{align}
    \mathrm{ACC}_K(t')
    = \frac{1}{N_\mathrm{Test}}\sum_{n=1}^{N_\mathrm{Test}}
    \mathbbm{1}\!\left\{{m^{(n)}_{t'}}^{*} \in \mathcal{Q}_K\big(\hat{\mathbf{y}}_{t'}^{(n)}\big)\right\},
    \end{align}
    where $\mathcal{Q}_K(\cdot)$ returns the $K$ beams with the highest predicted probabilities from
    the score vector $\hat{\mathbf{y}}_{t'}^{(n)}$ for sample $n$.  
    We report results for $K\in\{1,3\}$; $\mathrm{ACC}_K$ follows the common practice of scheduling
    a short candidate beam list and is less sensitive to small angular label noise \cite{3gpp,industry}.
    
    \subsection{Results and Discussion}
    \subsubsection{Training Dynamics and Convergence}

\begin{figure}[t]
    \captionsetup{font=footnotesize}
    \begin{center}
    \includegraphics[width=\linewidth]{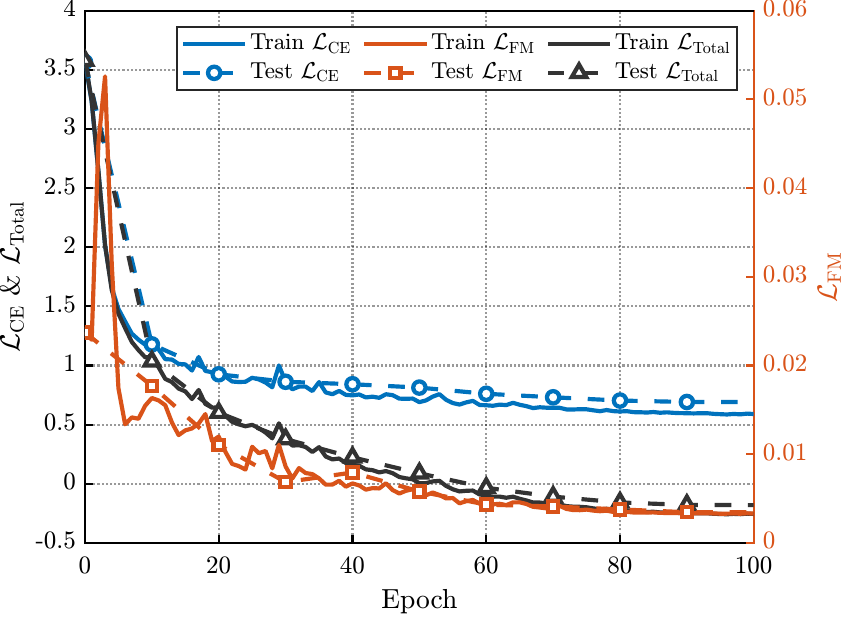}
    \end{center}
    \caption{Training and testing loss curves.}
    \label{fig:loss}
    \end{figure}

    Fig.~\ref{fig:loss} shows all objectives dropping sharply in the first 20 epochs and then gradually stabilizing after epoch 50. The steady drop of $\mathcal{L}_\mathrm{FM}$ indicates that the learned vector field successfully matches the \textcolor{black}{target vectors of local beam evolution}. Both $\mathcal{L}_\mathrm{CE}$ and $\mathcal{L}_\mathrm{Total}$ show similar decreasing trends. The absence of late-stage oscillation or divergence confirms a stable training process.

    \subsubsection{Beam Prediction Performance}

    \begin{figure}[t]
    \captionsetup{font=footnotesize}
    \begin{center}
    \includegraphics[width=\linewidth]{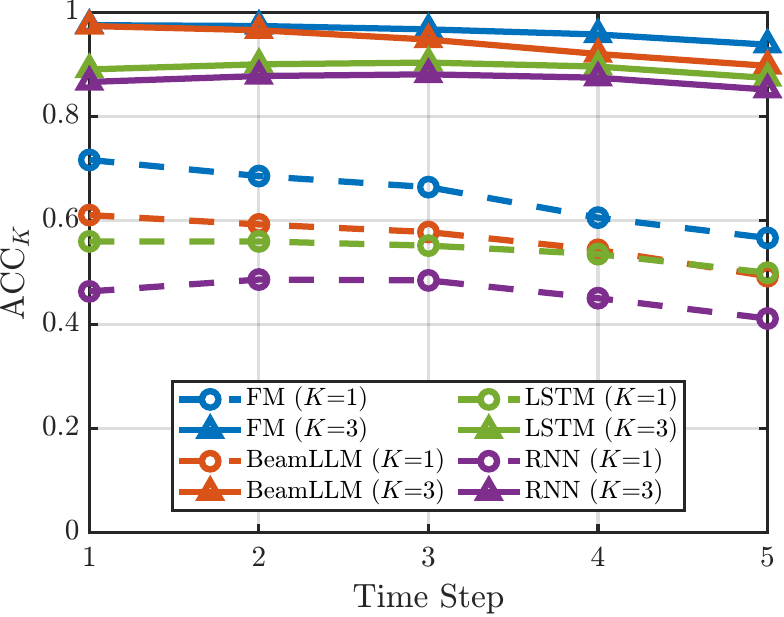}
    \end{center}
    \caption{$\mathrm{ACC}_K$ performance of the proposed method compared with several baselines under configuration A ($T_\mathrm{Hist} = 8$ and $T_\mathrm{Pred} = 5$).}
    \label{fig:std}
    \end{figure}

    \begin{figure}[t]
    \captionsetup{font=footnotesize}
    \begin{center}
    \includegraphics[width=\linewidth]{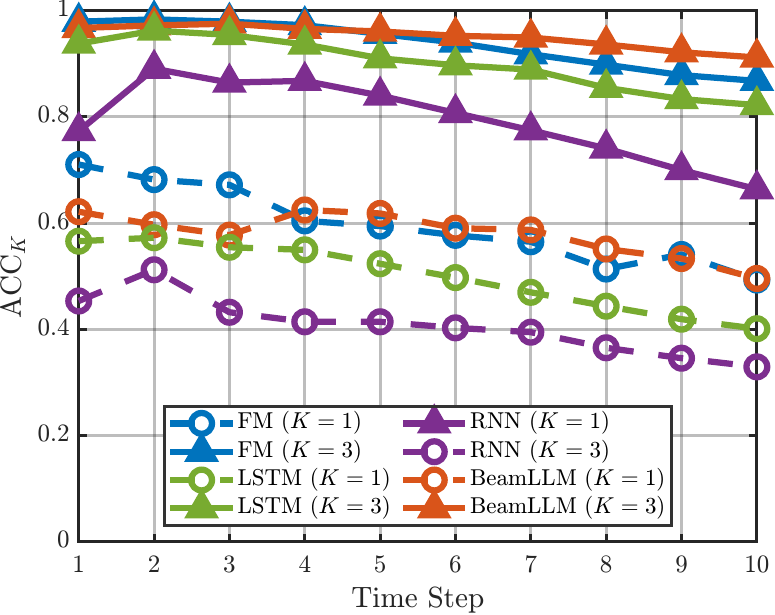}
    \end{center}
    \caption{$\mathrm{ACC}_K$ performance of the proposed method compared with several baselines under configuration B ($T_\mathrm{Hist} = 3$ and $T_\mathrm{Pred} = 10$).}
    \label{fig:long}
    \end{figure}

    Under configuration A, Fig.~\ref{fig:std} compares $\mathrm{ACC}_1$ and $\mathrm{ACC}_3$ over prediction steps. FM yields the best performance throughout. $\mathrm{ACC}_1$ for all models gradually decreases over time, but FM consistently keeps the highest values. $\mathrm{ACC}_3$ remains relatively stable with slight degradation, where FM also stays superior and its margin grows mildly with time. Among baselines, BeamLLM is closest to FM, followed by LSTM, while RNN performs worst in both metrics.

    \begin{figure}[t]
        \centering
        \captionsetup{font=footnotesize}
        \includegraphics[width=\linewidth]{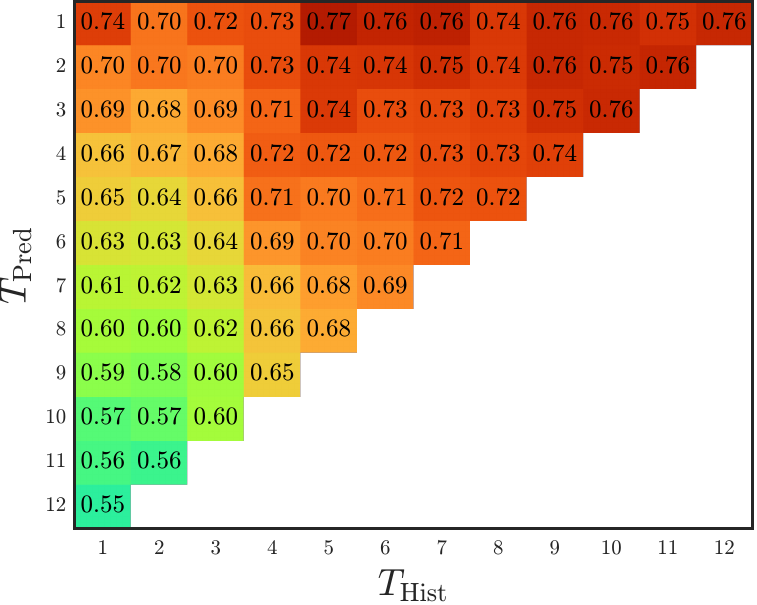}
        \caption{Heatmap of $\mathrm{ACC}_1$ under varying $T_{\mathrm{Hist}}$ and $T_{\mathrm{Pred}}$.}
        \label{fig:top1_heatmap}
    \end{figure}

    \textcolor{black}{Turning to the longer prediction horizon in configuration B,} Fig. \ref{fig:long} shows the model performance across prediction steps.
    BeamLLM demonstrates the strongest long-term robustness: although it is not the top performer in the early stages, it consistently improves and eventually becomes the best-performing model, maintaining the most stable performance over time.
    FM attains strong early accuracy but degrades in later steps, particularly for $\mathrm{ACC}_1$.
    LSTM presents \textcolor{black}{a smoother but consistently lower performance trend}, while RNN performs the worst throughout the horizon with the most substantial deterioration.
    Overall, BeamLLM is preferable for long-term prediction due to its higher and more stable late-step accuracy.

    \textcolor{black}{To examine whether the above trends hold beyond the two representative configurations,} Fig.~\ref{fig:top1_heatmap} illustrates the joint impact of $T_{\mathrm{Hist}}$ and $T_{\mathrm{Pred}}$ on average $\mathrm{ACC}_1$. Overall, the prediction performance is positively correlated with $T_{\mathrm{Hist}}$ and negatively correlated with $T_{\mathrm{Pred}}$. Specifically, increasing $T_{\mathrm{Hist}}$ effectively improves prediction accuracy, as longer visual sequences provide the model with richer spatial-temporal context, user movement trajectories, and environmental dynamics. Conversely, as the prediction step $T_{\mathrm{Pred}}$ increases, the accuracy degrades. This degradation occurs because the uncertainty of the channel environment accumulates over time, making prediction much more difficult. \textcolor{black}{However}, longer visual sequences \textcolor{black}{also} increase the computational overhead and delay during image processing and model inference. Therefore, in practical vision-aided communication systems, $T_{\mathrm{Hist}}$ must be carefully chosen based on service reliability requirements to achieve an optimal engineering trade-off between beam prediction accuracy and the computational cost.
    
    \subsubsection{Ablation Studies}

    Table~\ref{tab:ablation} reports the average $\mathrm{ACC}_K$ under configurations A and B. The full model achieves the best performance on all metrics, showing the overall effectiveness of the proposed design. Removing $\mathcal{L}_{\mathrm{FM}}$ causes the largest performance drop in both configurations, especially for long-term prediction, indicating that learning the local evolution of beam power distributions is essential for accurate beam prediction. Replacing the condition encoder $f_{\mathrm{cond}}(\cdot)$ with an LSTM leads to only a small decrease in accuracy, suggesting that the model is not highly sensitive to the specific choice of sequential encoder. A similar trend can be observed when using an RNN-based condition encoder, although the performance drop is slightly larger in some cases. Finally, the rectified flow variant also performs worse than the full model in all settings \cite{rectified_flow}, which further confirms the advantage of the proposed segment-wise modeling. Overall, these results demonstrate the effectiveness of the segment-wise FM framework for improving prediction accuracy and adapting to complex beam dynamics.  
    
    \begin{table}[t]
        \footnotesize
        \centering
        \caption{Ablation study on average $\mathrm{ACC}_K$. The best result for each entry is highlighted in bold and underscored.}
        \label{tab:ablation}
        \setlength{\tabcolsep}{2pt}
        \renewcommand{\arraystretch}{1.25}
        
        \begin{tabularx}{\linewidth}{c|*{10}{>{\centering\arraybackslash}X}} 
        \hline 
        \textbf{Metrics} &
        \multicolumn{2}{c|}{\textbf{Base}} &
        \multicolumn{2}{c|}{\textbf{w/o} $\mathcal{L}_{\mathrm{FM}}$} &
        \multicolumn{2}{c|}{\makecell{\textbf{LSTM} \\ \textbf{cond.}}} &
        \multicolumn{2}{c|}{\makecell{\textbf{RNN} \\ \textbf{cond.}}} &
        \multicolumn{2}{c}{\makecell{\textbf{Rectified} \\ \textbf{Flow}}} \\
        \hline
        $K$ & 1 & 3 & 1 & 3 & 1 & 3 & 1 & 3 & 1 & 3 \\
        \cline{2-11}
        \hline
        cfg. A & \underline{$\mathbf{0.72}$} & \underline{$\mathbf{0.99}$} & $0.41$ & $0.80$ & $0.70$ & $0.98$ & $0.65$ & $0.96$ & $0.68$ & $0.96$ \\
        cfg. B & \underline{$\mathbf{0.60}$} & \underline{$\mathbf{0.94}$} & $0.24$ & $0.56$ & $0.54$ & $0.91$ & $0.54$ & $0.90$ & $0.57$ & $0.92$ \\
        \hline
        \end{tabularx}
    \end{table}

    \subsubsection{Complexity Analysis}
    Inference experiments are conducted on an NVIDIA Tesla T4 GPU and Intel Xeon CPU available on Google Colab. Table~\ref{tab:complexity} summarizes model complexity and latency. The proposed FM model achieves comparable computational cost to lightweight baselines while offering higher accuracy, making it well-suited for edge deployment. In comparison,  BeamLLM incurs significantly higher CPU-side delay. Therefore, all evaluated models except BeamLLM on CPU are practically deployable, while FM-based model stands out as a more balanced option for performance-critical edge inference scenarios.

    \begin{table}[t]
    \centering
    \caption{Comparison of the network parameters and the inference cost per test sample.}
    \small
    \setlength{\tabcolsep}{6pt}
    \renewcommand{\arraystretch}{1.2}
    \begin{tabularx}{\linewidth}{l X c c}
        \toprule
        \textbf{Model}
        & \#\textbf{ Parameters}
        & \makecell{\textbf{Inf. Time}\\\textbf{(GPU, sec)}}
        & \makecell{\textbf{Inf. Time}\\\textbf{(CPU, sec)}} \\
        \midrule
        RNN           & $30$ K     & $2.0\times 10^{-4}$    & $1.9\times 10^{-4}$ \\
        LSTM          & $104$ K    & $2.2\times 10^{-4}$    & $2.8\times 10^{-4}$ \\
        FM & $195$ K     & $3.2\times 10^{-4}$    & $4.7\times 10^{-4}$ \\
        BeamLLM       & $178$ M & $2.2\times 10^{-3}$ & $1.31$ \\
        \bottomrule
    \end{tabularx}
    \begin{tablenotes}
        \footnotesize
        \item \textit{Note: }The object detector’s parameters and runtime are excluded to focus on the beam predictor’s complexity, since its visual features are shared and reusable for other tasks (e.g., vehicle statistics).

    \end{tablenotes}
    \label{tab:complexity}
    \end{table}

\section{Conclusion}
\label{chapt5}

We proposed a segment-wise FM framework for vision-aided mmWave beam prediction that models continuous beam dynamics for fast and stable inference. The proposed method achieves higher prediction accuracy than traditional sequence models and competitive performance relative to BeamLLM, while requiring substantially lower computational complexity. This makes it a promising lightweight predictor for edge-side beam prediction.

Future work will extend the framework to more general and challenging settings, including multi-user scenarios, more diverse traffic densities, adverse weather conditions, and different sensing configurations. Beyond beam prediction, we aim to explore FM-based modeling for other wireless tasks, such as proactive channel prediction in high-mobility links and traffic or network load forecasting.
\section*{Acknowledgment}
\vspace*{-1mm}
{\small This work was supported by the National Research Foundation of Korea (NRF) grant funded by the Korea government (MSIT) (RS-2025-00517140).}
\vspace*{-1mm}

\balance 
\bibliographystyle{IEEEtran}
\bibliography{ref}
\end{document}